\documentclass[reprint, superscriptaddress, nofootinbib,aps,prl]{revtex4-1}
\pdfoutput=1
\usepackage{mathtools, amsfonts, amsthm, latexsym, amssymb,dsfont}
\usepackage[T1]{fontenc}
\usepackage{microtype}
\usepackage{hyperref}
\hypersetup{colorlinks=true, citecolor=blue, urlcolor=blue}
\usepackage{bm,bbm}
\usepackage{graphicx}
\usepackage{dcolumn}
\usepackage{parskip}
\usepackage{xcolor}
\usepackage{cancel}

\usepackage{blkarray}

\def\CC{{\cal C}}
\def\CD{{\cal D}}

\def\CO{{\cal O}}

\def\b0{\bm{0}_\perp}

\usepackage{color}

\def\slO{\mathsf{O}}

\usepackage{centernot}
\usepackage{mathtools}
\usepackage{stmaryrd}

\makeatletter
\newcommand{\xMapsto}[2][]{\ext@arrow 0599{\Mapstofill@}{#1}{#2}}
\def\Mapstofill@{\arrowfill@{\Mapstochar\Relbar}\Relbar\Rightarrow}
\makeatother

\DeclareFontFamily{U}{mathx}{\hyphenchar\font45}
\DeclareFontShape{U}{mathx}{m}{n}{
      <5> <6> <7> <8> <9> <10>
      <10.95> <12> <14.4> <17.28> <20.74> <24.88>
      mathx10
      }{}
\DeclareSymbolFont{mathx}{U}{mathx}{m}{n}
\DeclareFontSubstitution{U}{mathx}{m}{n}
\DeclareMathAccent{\widecheck}{0}{mathx}{"71}

\begin{document}
\title{Method of images in defect conformal field theories}

\author{Tatsuma Nishioka}
\affiliation{
Department of Physics, Osaka University,\\
Machikaneyama-Cho 1-1, Toyonaka 560-0043, Japan
}
\author{Yoshitaka Okuyama}
\affiliation{
Department of Physics, Osaka University,\\
Machikaneyama-Cho 1-1, Toyonaka 560-0043, Japan
}
\affiliation{Department of Physics, Faculty of Science,
The University of Tokyo,\\
Bunkyo-Ku, Tokyo 113-0033, Japan}
\author{Soichiro Shimamori}
\affiliation{
Department of Physics, Osaka University,\\
Machikaneyama-Cho 1-1, Toyonaka 560-0043, Japan
}

\date{\today}

\begin{abstract}
We propose a prescription for describing correlation functions in higher-dimensional defect conformal field theories (DCFTs) by those in ancillary conformal field theories (CFTs) without defects, which is a vast generalization of the image method in two-dimensional boundary CFTs.
A correlation function of $n$ operators inserted away from a defect in a DCFT is represented by a correlation function of $2n$ operators in the ancillary CFT, each pair of which is placed symmetrically with respect to the defect.
We establish the correspondence by matching the constraints on correlation functions imposed by conformal symmetry on both sides.
Our method has the potential to shed light on new aspects of DCFTs from the viewpoint of conventional CFTs.
\end{abstract}
\maketitle

The method of images is one of the most powerful mathematical techniques for solving differential equations under a boundary condition by extending the domain of a function across the boundary that mirrors the images of the original domain to the outside.
It is a highly versatile method applicable to various situations, ranging from classical electromagnetism to condensed matter physics to string theory.
The most familiar and intuitive physical application one encounters is the calculation of the electric field of a charged particle in the presence of a conducting surface.
Another well-known example in condensed matter physics relates to topological insulators.
In the presence of the $\mathrm{U}(1)$ topological term, when an electrically charged particle is placed near the gapless surface of a topological insulator, a magnetic monopole appears as a mirror image of the electric charge \cite{2009Sci...323.1184Q}.

Interestingly, a similar idea of having mirror images was proposed by Cardy \cite{Cardy:1984bb} in two-dimensional boundary conformal field theory (BCFT${}_{2}$), and applied to reduce the correlation functions of local operators to those of the local operators and their mirror images in a two-dimensional conformal field theory (CFT${}_2$) without boundary.
This is known as the ``doubling trick'' as it doubles the number of operators inside the correlator.
The validity of this trick follows straightforwardly from the fact that both the correlation functions in BCFT$_2$ and the corresponding CFT$_2$ satisfy the same conformal Ward identities.
This perspective allows us to investigate BCFT$_2$ by leveraging various techniques and results in CFT$_2$. Prominent examples include quantum impurity problems (the Kondo effect) \cite{LudwigAffleck:1991,Affleck:1995ge}, spin-spin correlation functions in the Ising model \cite{Oshikawa:1996dj}, open string theories and D-branes \cite{Polchinski:1998rq,Recknagel:2013uja}, and quantum quenches in CFT$_2$ \cite{Calabrese:2006rx,Calabrese:2016xau}.

More generally, BCFT falls into a subclass of defect CFTs (DCFTs) with extended objects called ``defects'' in addition to local operators.
From the DCFT point of view, the boundary of BCFT in $d$ dimensions can be regarded as a $(d-1)$-dimensional defect.
DCFTs with lower-dimensional defects have attracted attention and been extensively investigated in recent works; boundary and defect conformal bootstrap \cite{Liendo:2012hy,Gaiotto:2013nva,Antunes:2021qpy}, Lorentzian inversion formula for two-point functions in DCFT \cite{Liendo:2019jpu}, 
classification of defect central charges \cite{Herzog:2015ioa,Herzog:2017kkj,Herzog:2017xha,FarajiAstaneh:2021foi,Jensen:2018rxu,Chalabi:2021jud}, defect $\CC$-theorems \cite{Kobayashi:2018lil,Jensen:2015swa,Cuomo:2021rkm,Wang:2021mdq}, integrable structures \cite{Giombi:2018hsx,Giombi:2018qox},
to name a few.
Further developments of DCFT in various fields may be found in \cite{proceeding:2018} and the references therein.

\begin{figure}
    \includegraphics[width=4.5cm]{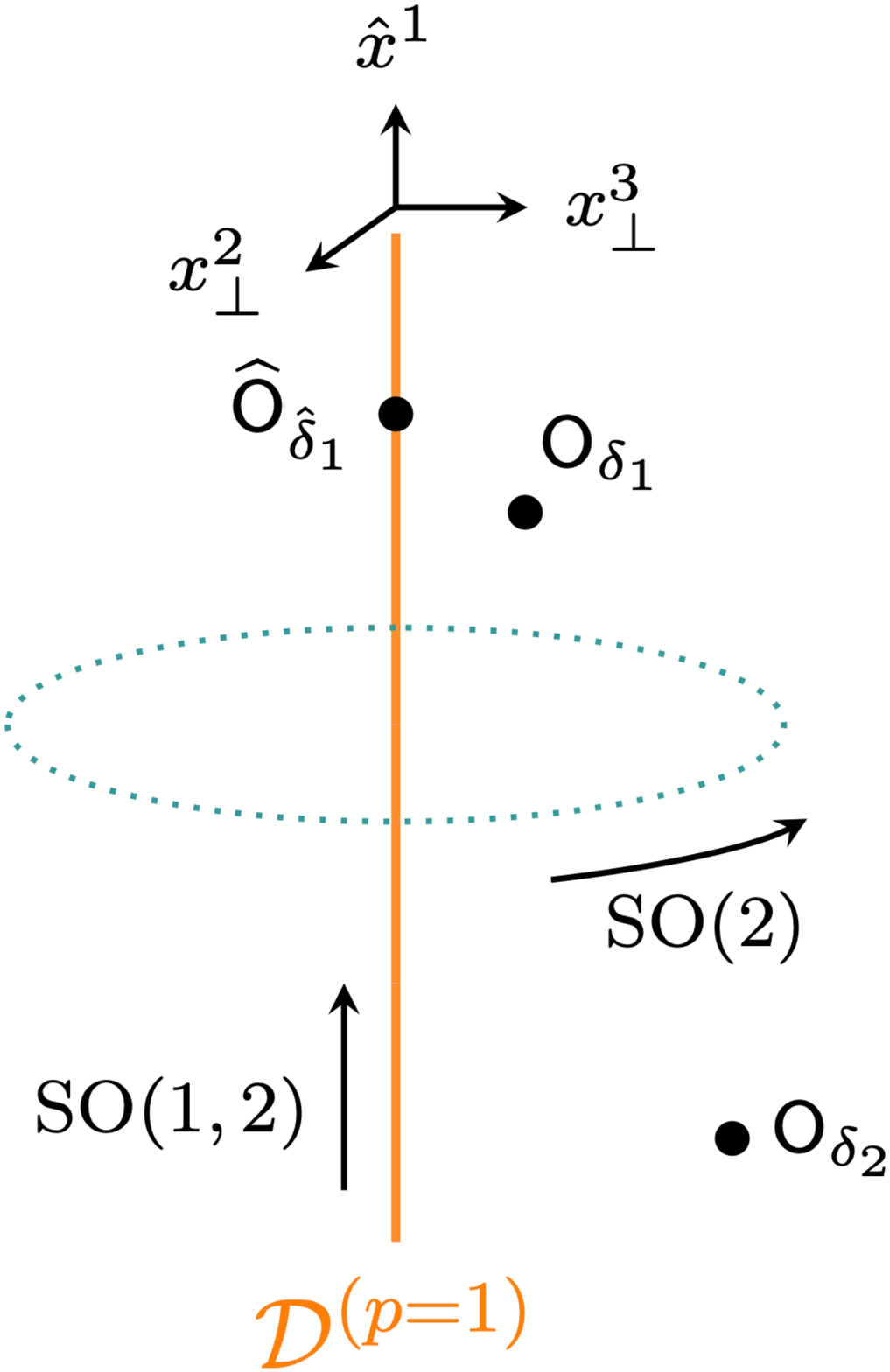}
 	\caption{\label{fig:residual symmetry}Illustrated is a line defect 
    $\CD^{(p=1)}$ located at $x_\perp^2=x_\perp^3=0$ in three-dimensional spacetime $\mathbb{R}^{d=3}$. In this case, the full conformal group $\mathrm{SO}(1,4)$ breaks into the product of the conformal group $\mathrm{SO}(1,2)$ and the rotation group $\mathrm{SO}(2)$.
}
\end{figure}

Adding an arbitrary defect to CFT typically breaks the entire conformal symmetry $\mathrm{SO}(1,d+1)$, while DCFT allows a class of defects (conformal defects of planar or spherical shape) that preserve a part of conformal symmetry.\footnote{Throughout this paper, we consider DCFTs in Euclidean spacetime.} The residual symmetry group of DCFT with a $p$-dimensional defect $\CD^{(p)}$ is $\mathrm{SO}(1,p+1)\times \mathrm{SO}(d-p)$ and is called the defect conformal group. The $\mathrm{SO}(1,p+1)$ and $\mathrm{SO}(d-p)$ groups correspond to the conformal and rotational groups that act on and around the defect, respectively.
Without loss of generality, we place a $p$-dimensional planer defect at $x^\mu=0$ for $\mu=p+1,\cdots d$ and decompose the $d$-dimensional coordinates $x^\mu$ into parallel and transverse directions to the defect as $x^\mu=(\hat{x}^a,x_\perp^i)$ with $a=1,\cdots, p$ and $i=p+1,\cdots, d$. As a simplest example, we illustrate a line defect ($p=1$) in a three-dimensional space $\mathbb{R}^{d=3}$ in Fig.\,\ref{fig:residual symmetry}.

In DCFT, there are two types of operators: {\it bulk local operators and defect local operators}.
In particular, a bulk scalar operator $\slO_{\delta}$ with conformal dimension $\delta$ is located away from the defect, while a defect local scalar operator $\widehat{\slO}_{\hat{\delta}}$ with conformal dimension $\hat{\delta}$ lives on the defect and represents localized excitations there.
Although a few initial studies of the correlation functions of bulk and defect local operators have been carried out in \cite{Billo:2016cpy,Gadde:2016fbj,Kobayashi:2018okw,Guha:2018snh}, these methods are rather involved and of less practical use so far (see, e.g., \cite{Herzog:2020bqw} for the current status).

In this paper, we propose a novel generalization of the method of images in DCFT, which translates correlation functions of DCFT into those of CFT.\footnote{The method of images was foreseen previously in the study of a line defect in the $(2+1)$-dimensional Ising model \cite{ko1985energy}.
This correspondence was also hinted by other recent researches in BCFT (e.g., \cite{Herzog:2017xha,Bissi:2018mcq,Dey:2020jlc}).
We thank J.\,H.\,H.\,Perk and A.\,S\"oderberg for informing us of their relevant works.} Our prescription provides us a clear-cut way to write down DCFT correlators without resorting to specialized techniques, and lowers the substantial barriers to investigating unexplored structures of defects in CFT.

We first focus on scalar primaries for simplicity and state our prescription. Then, we move on to the spinning case.

Our statement for scalar primaries is as follows: 

{\it The correlation function of $n$ bulk scalars $\slO_{\delta_\alpha},\alpha=1,\cdots, n$ and $m$ defect local scalars $\widehat{\slO}_{\hat{\delta}_{\hat{\alpha}}},\hat{\alpha}=1,\cdots, m$ in DCFT is equivalent to the correlation function of $n$-pairs of local scalars $\CO_{\delta_\alpha/2}$ and $m$ local scalars $\CO_{\hat{\delta}_{\hat{\alpha}}}$ in an ancillary CFT by the following relation:}
{\footnotesize\begin{align}\label{statement}
\begin{aligned}
        & \left\langle\,\prod_{\alpha=1}^{n}\,\slO_{\delta_{\alpha}} (x_{\alpha}) \,\prod_{\hat{\alpha}=1}^{m}\,\widehat{\slO}_{\hat{\delta}_{\hat{\alpha}}} (y_{\hat{\alpha}})\, \right\rangle_{\text{DCFT}}\\
   &\qquad\approx  \left\langle\,\prod_{\alpha=1}^{n}\,\left[\CO_{\delta_{\alpha}/2}(x_{\alpha})\,\CO_{\delta_{\alpha}/2} (\bar{x}_{\alpha})\right]\, \prod_{\hat{\alpha}=1}^{m}\,\CO_{\hat{\delta}_{\hat{\alpha}}} (y_{\hat{\alpha}})\,\right\rangle_{\text{CFT}}\ ,
\end{aligned} \tag{$\mathbf{S}$}
\end{align}}
{\it where $\approx$ means that both sides satisfy the same differential equations dictated by conformal symmetry. The coordinate $\bar{x}$ stands for the antipodal point of $x$ along the transverse direction to the boundary/defect: $\bar{x}^{\mu}= (\hat x^{a}, -x^{i}_\perp)$, whereas $y$ is the coordinate on the defect, $y^\mu =(\hat{y}^a,y_\perp^i=0)$.}\footnote{We consider a theory with defects invariant under antipodal reflection $x\mapsto \bar{x}$.}

We emphasize that the ancillary CFT considered here is neither the bulk part of the DCFT nor necessarily unitary in general. 
To clarify this point, let us take as an example a line defect in a free scalar theory in four-dimensional spacetime \cite{Kapustin:2005py,Billo:2016cpy}. 
The free scalar field is a bulk scalar primary operator of dimension $\delta=1$, so the conformal dimension of the corresponding auxiliary scalar on the CFT side becomes $\delta/2=1/2$, which is clearly below the unitarity bound in four-dimensional CFT \cite{Poland:2018epd}.
The ancillary CFT correlators  $\langle\,\cdots\,\rangle_{\text{CFT}}$ considered in the statement \eqref{statement} serve just as conformally invariant structures and should not be regarded as physical objects.
To make manifest the auxiliary role of the CFT, we use different fonts to denote the operators in DCFT and the associated CFT.

\begin{figure}
	\centering
    \includegraphics[width=9cm]{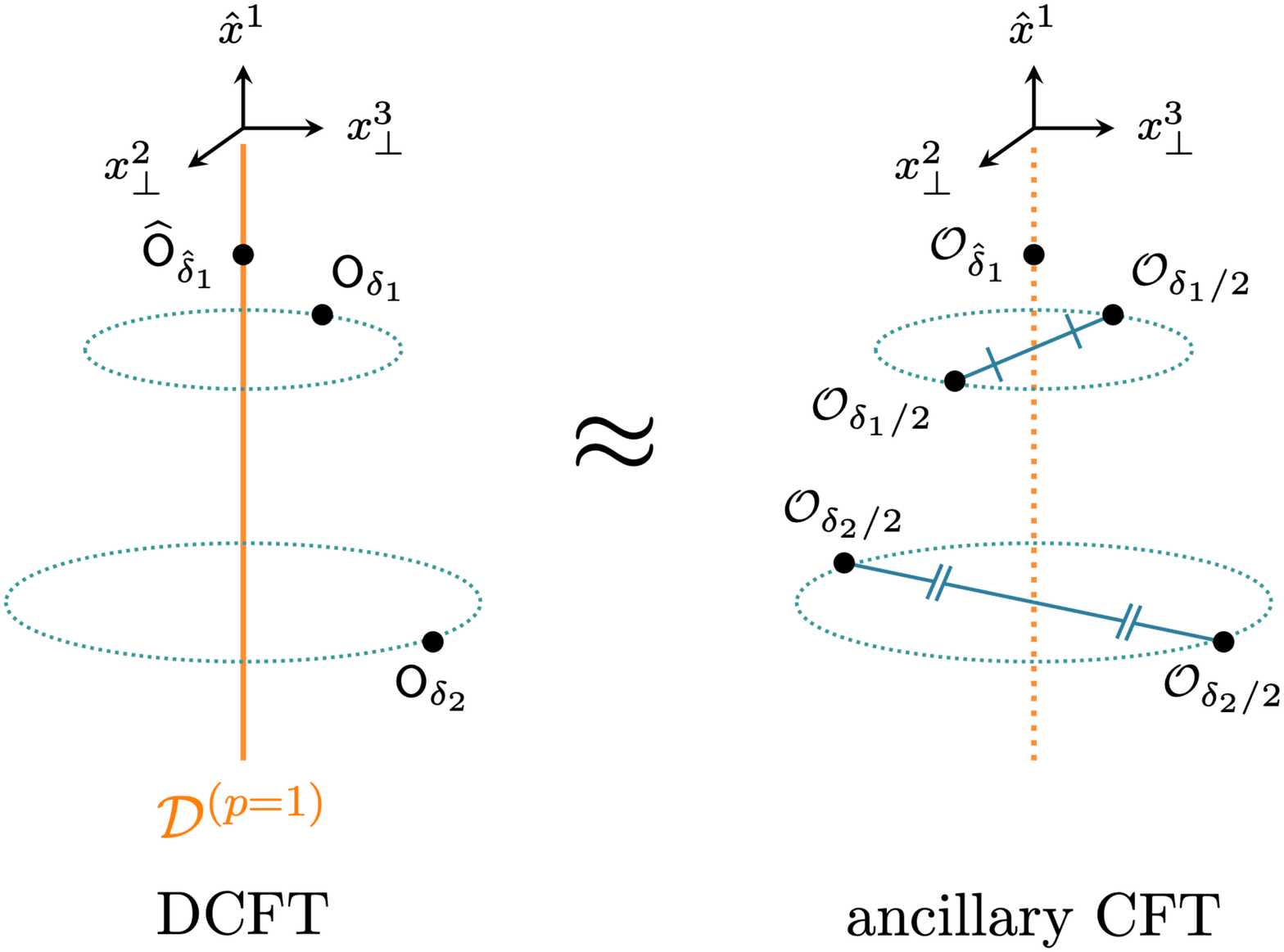}
	\caption{The method of images \eqref{statement} for $n=2,m=1$ in three-dimensional spacetime with a line defect. The bulk-bulk-defect three-point function in DCFT [Left] behaves in the same way as the five-point conformally invariant structure [Right]. Each bulk scalar primary in DCFT is associated with a pair of auxiliary scalar primaries placed in the mirror-symmetric configuration with respect to the position of the defect in an ancillary CFT.}
	\label{fig:defect}
\end{figure}

\begin{figure}
	\centering
    \includegraphics[width=9cm]{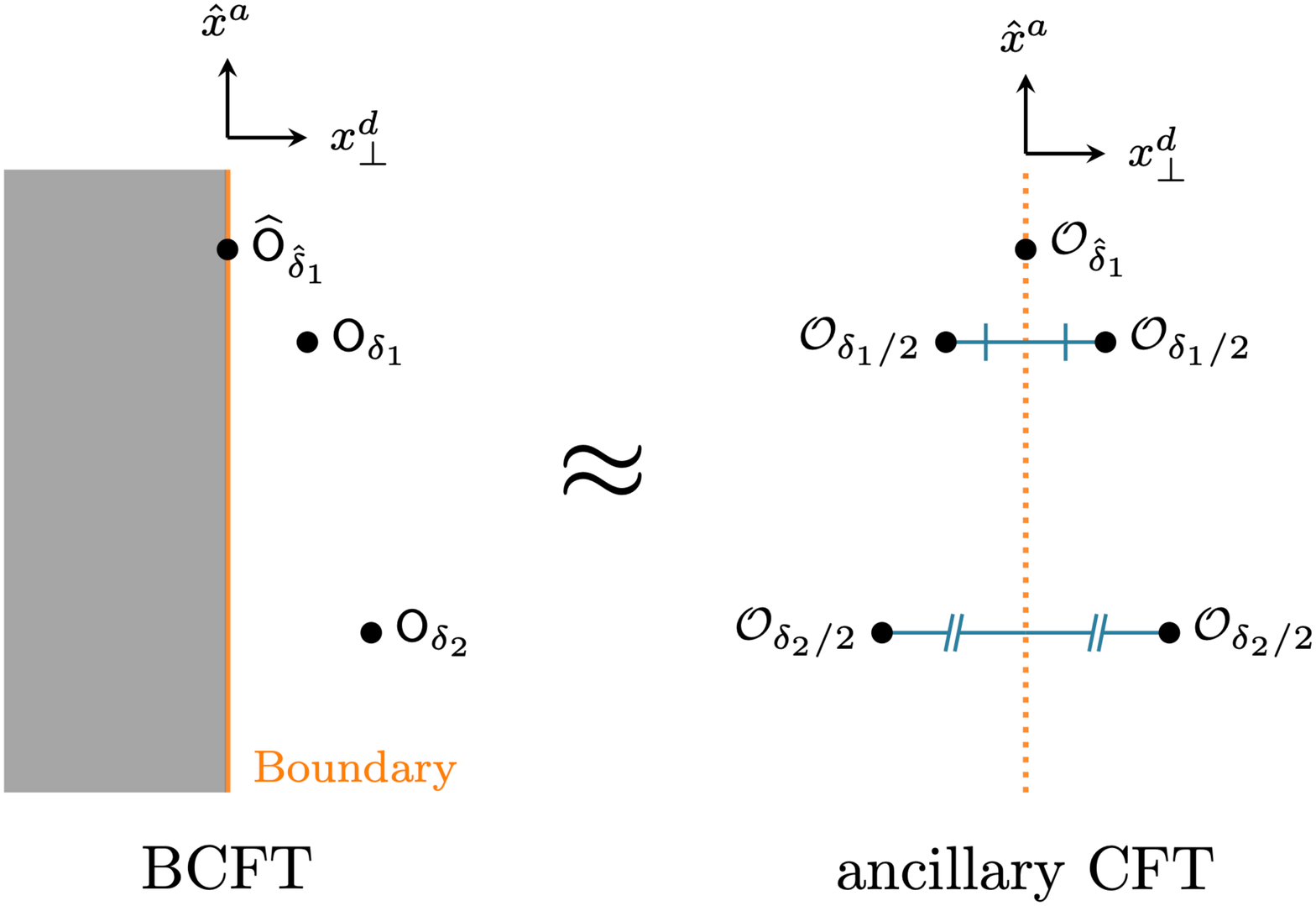}
	\caption{The method of images \eqref{statement} for BCFT with two bulk and one boundary operators. The BCFT correlator on upper half-space $x_{\perp}^d\geq 0$ [Left] is equivalent to the corresponding ancillary CFT correlator on the full space [Right].}
\label{fig:boundary}
\end{figure}

Figure \ref{fig:defect} and \ref{fig:boundary} show two specific cases of the statement \eqref{statement}; a line defect $(p=1)$ in three dimensions $(d=3)$ and $d$-dimensional BCFT $(p=d-1)$, respectively.

We now present a few working examples of the correspondence \eqref{statement} in order.

\bigskip
\paragraph{Bulk one-point function.}
The one-point function of a bulk local operator in DCFT is uniquely fixed by the conformal symmetry to be
\begin{align}\label{eq:1pt_DCFT}
    \langle\, \slO_\delta (x)\, \rangle_{\text{DCFT}}
    =
    \frac{a_{\delta}}{|x_\perp|^\delta} \ ,
\end{align}
where $|x_\perp|^{2}\equiv x^{i}_{\perp}x_{\perp,i}$ and $a_{\delta}$ is a constant whose value depends on a model of DCFT.
On the other hand, the two-point function of bulk operators inserted at $x^\mu = (\hat x^a, x^i_\perp)$ and the antipodal point $\bar x^\mu = (\hat x^a, -x^i_\perp)$ is
\begin{align}\label{eq:2pt CFT mirror}
    \langle\,\CO_{\delta/2} \left(x\right)\, \CO_{\delta/2} \left(\bar x\right)\,\rangle_{\text{CFT}}
    =
    \frac{C_{\delta/2}}{|x - \bar x|^{\delta}}
    =
    \frac{C_{\delta/2}}{2^{\delta}\,|x_\perp|^{\delta}}\ ,
\end{align}
where $C_{\delta/2}$ is a constant.
It reproduces the DCFT one-point function \eqref{eq:1pt_DCFT} up to a coefficient.

\bigskip
\paragraph{Bulk-to-defect two-point function.}
Next, we consider the two-point function of a bulk operator at $x^\mu$ and a defect local operator at $y^\mu=(\hat y^a, 0)$:
\begin{align}\label{eq:2pt_DCFT}
     \langle\, \slO_\delta (x)\,\widehat\slO_{\hat{\delta}} (y)\, \rangle_{\text{DCFT}}
    =
    \frac{b_{\delta,\hat{\delta}}}{|x_\perp|^{\delta-\hat{\delta}}\left(|\hat x - \hat y|^2 + |x_\perp|^2 \right)^{\hat{\delta}}} \ ,
\end{align}
where $|\hat{x}|^{2}\equiv \hat{x}^{a}\hat{x}_{a}$ and $b_{\delta,\hat{\delta}}$ is a model-dependent constant.
Applying the method of images yields the following three-point function in CFT:
\begin{align}\label{eq:3pt aCFT}
    \begin{aligned}
     \langle\,\CO_{\delta/2} (x)\,&\CO_{\delta/2} (\bar x)\,\CO_{\hat{\delta}} (y)\,\rangle_{\text{CFT}}\\
            &=
            \frac{C_{\delta/2,\delta/2,\hat{\delta}}}{|x-y|^{\hat{\delta}}\,|y-\bar{x}|^{\hat{\delta}}\,|x-\bar x|^{\delta - \hat{\delta}}} \\
            &=
            \frac{C_{\delta/2,\delta/2,\hat{\delta}}}{2^{\delta - \hat{\delta}}\,|x_\perp|^{\delta-\hat{\delta}}\left(|\hat x - \hat y|^2 + |x_\perp|^2 \right)^{\hat{\delta}}} \ ,
    \end{aligned}
\end{align}
with an undetermined constant $C_{\delta/2,\delta/2,\hat{\delta}}$.
This correlator correctly reproduces the bulk-to-defect two-point function \eqref{eq:2pt_DCFT} up to a coefficient as expected. 

\bigskip
\paragraph{Outline of the proof of the image method.}
Conformal correlators are strongly constrained by differential equations associated with conformal symmetry. These equations are often called {\it the conformal Ward identities}. Therefore, it is enough to show that both hand sides of \eqref{statement} satisfy the same differential equations related to common symmetries: the conformal symmetry parallel to the defect and the rotational symmetry around the defect. Here we illustrate this strategy by taking the bulk one-point function as a simplest example. We will check that both $\langle\, \slO_\delta (x)\, \rangle_{\text{DCFT}}$ and $\langle\,\CO_{\Delta_1} (x_1)\, \CO_{\Delta_2} (x_2)\,\rangle_{\text{CFT}}$ satisfy the same differential equations inherited from the residual symmetry $\mathrm{SO}(1,p+1)\times \mathrm{SO}(d-p)$ if we place two primaries mirror-symmetrically in the CFT side, namely, $x_1=x$, $x_2=\bar{x}$, and $\Delta_1=\Delta_2=\delta/2$.
\paragraph{-- Translations in the parallel directions.}
The conformal Ward identities in DCFT associated with the translations parallel to the defect are
\begin{align}\label{1pt_DCFT_Pa}
    \frac{\partial}{\partial \hat x^{a}}\,\langle\,\slO_\delta (x)\,\rangle_{\text{DCFT}} =  0    \ ,
\end{align}
    while on the CFT side the conformal Ward identities along the $x^a$ direction are
    \begin{align}\label{2pt_CFT_Pa}
    \begin{aligned}
     &\left\langle\,\frac{\partial}{\partial x_{1}^a}\,\CO_{\Delta_{1}} (x_1)\, \CO_{\Delta_2} (x_2)\, \right\rangle_{\text{CFT}} \\
     &\qquad\qquad
     +\left\langle\,\CO_{\Delta_{1}} (x_1)\,\frac{\partial}{\partial x_{2}^a}\,\CO_{\Delta_2} (x_2)\,\right\rangle_{\text{CFT}}=0 \ .
     \end{aligned}
    \end{align}
If we take mirror-symmetric configuration ($x_1=x$, $x_2=\bar{x}$ and $\Delta_1=\Delta_2=\delta/2$), the differential equations \eqref{2pt_CFT_Pa} reduce to
\begin{align}
      \frac{\partial}{\partial x^{a}} \langle\,\CO_{\delta/2} \left(x\right)\, \CO_{\delta/2} \left(\bar x\right)\,\rangle_{\text{CFT}} =0\ ,
\end{align}
which correctly reproduces \eqref{1pt_DCFT_Pa}.   
\paragraph{-- Dilatation.}
On the DCFT side, the dilation symmetry yields the conformal Ward identity
 \begin{align}\label{1pt_DCFT_D}
    \left(x^{\mu}\frac{\partial}{\partial x^{\mu}}+\delta\right)\,\langle\,\slO_\delta (x)\,\rangle_{\text{DCFT}}  = 0  \ ,
\end{align}
 while in CFT we have
 \small
 \begin{align}\label{2pt_CFT_D}
    \begin{aligned}
    & \left\langle\, \left(x_1^{\mu}\,\frac{\partial}{\partial x_1^{\mu}}+\Delta_1\right)\,\CO_{\Delta_{1}} (x_1)\,\CO_{\Delta_2} (x_2)\,\right\rangle_\text{CFT}\\
    & \quad+\left\langle\,\CO_{\Delta_{1}}(x_2)\, \left(x_2^{\mu}\,\frac{\partial}{\partial x_2^{\mu}}+\Delta_2\right)\, \CO_{\Delta_2} (x_2)\right\rangle_{\text{CFT}}=0\ .
     \end{aligned}
    \end{align}
\normalsize
By locating $x_1, x_2$ at the mirror-symmetric configuration and setting $\Delta_1=\Delta_2=\delta/2$, \eqref{2pt_CFT_D} takes the same form as \eqref{1pt_DCFT_D}:
    \begin{align}
        \left(x^{\mu}\frac{\partial}{\partial x^{\mu}}+\delta\right)\,\langle\,\CO_{\delta/2} \left(x\right)\, \CO_{\delta/2} \left(\bar x\right)\,\rangle_{\text{CFT}} =0 \ .
    \end{align}
\paragraph{-- Special conformal transformations in the parallel directions.}
The conformal Ward identities for the special conformal transformations in DCFT are given by
\begin{align}\label{1pt_DCFT_Ka}
    \begin{aligned}
\left[2\hat{x}_{a}\left(x^{\mu}\frac{\partial}{\partial x^{\mu}}+\delta\right)-x^2\,\frac{\partial}{\partial \hat{x}^{a}}\right]\,\langle\,\slO_\delta (x)\,\rangle_{\text{DCFT}} = 0 \ , 
    \end{aligned}
    \end{align}
    while the conformal Ward identities for the special conformal transformations along the $x^a$ direction in CFT are
 \small\begin{align}\label{2pt_CFT_Ka}
    \begin{aligned}
     &\left\langle\, \left[2\hat{x}_{1,a}\left(x_1^{\mu}\frac{\partial}{\partial x_1^{\mu}}+\Delta_1\right)-x_1^2\frac{\partial}{\partial x_{1}^{a}}\right]\,\CO_{\Delta_{1}} (x_1)\, \CO_{\Delta_2} (x_2)\, \right\rangle_{\text{CFT}} 
      \\
    & +\left\langle\,\CO_{\Delta_{1}} (x_1)\, \left[2\hat{x}_{2,a}\left(x_2^{\mu}\frac{\partial}{\partial x_2^{\mu}}+\Delta_2\right)-x_2^2\,\frac{\partial}{\partial x_{2}^{a}}\right]\,\CO_{\Delta_2} (x_2)\right\rangle_{\text{CFT}}  \\
     & =0 \ .
     \end{aligned}
    \end{align}\normalsize
    In the mirror-symmetric configuration with $\Delta_1 = \Delta_2 = \delta/2$, \eqref{2pt_CFT_Ka} gives rise to the same differential equation as \eqref{1pt_DCFT_Ka}. \\
    Similarly, $\langle\,\slO_\delta (x)\,\rangle_{\text{DCFT}}$ and $\langle\,\CO_{\delta/2} \left(x\right)\, \CO_{\delta/2} \left(\bar x\right)\,\rangle_{\text{CFT}}$ satisfy the same differential equations for the parallel and transverse rotational symmetries.
    Thus, we have verified our statement \eqref{statement} for a bulk one-point function. 

By applying the same strategy as the one-point case to higher-point functions, $2n$-point CFT correlators can be shown to satisfy the same conformal Ward identities for $n$-point DCFT correlators when the operators are located in the mirror-symmetric configurations and their conformal dimensions are chosen appropriately.
We will provide the complete proof of \eqref{statement} to the upcoming paper \cite{long}, where we will use the embedding space formalism \cite{Costa:2011dw} to simplify the derivation by dealing with the conformal Ward identities more concisely.

\bigskip
\paragraph{Method of images for spinning primaries.}
The method of images for scalar operators takes a similar form to Cardy's doubling trick in two-dimensional BCFTs, where the holomorphic factorization allows to split a bulk local operator into the holomorphic and antiholomorphic parts. 
In higher-dimensional spacetime, the generalization of the image method for spinning operators is more intricate than the scalar case as the tensor structure of spinning operators is quite different from the two-dimensional case, and the holomorphic factorization no longer holds.
To circumvent the difficulties for dealing with symmetric traceless tensors, we use the encoding polynomial techniques \cite{Costa:2011mg}. 
Let us construct the encoding polynomial for a spin-$J$ symmetric traceless tensor $f_{\mu_1\cdots \mu_J}(x)$ by contracting its indices with a null polarization vector $z^\mu$ ($z^\mu z_\mu=0$):
\begin{align}
   f_J(x,z)\equiv z^{\mu_1}\cdots z^{\mu_J}\,f_{\mu_1\cdots \mu_J}(x)\ .
\end{align}
Then, we argue a kinematical correspondence between a spinning operator in DCFT and a pair of ancillary CFT operators located mirror-symmetrically against the defect with respect to the position $x$ as well as the polarization vector $z$: 
\small\begin{align}\label{statement spinning}
                       \begin{aligned}
                               & \left\langle\,\prod_{\alpha=1}^{n}\,\slO_{\delta_{\alpha},J_{\alpha}} (x_{\alpha},z_{\alpha}) \, \right\rangle_{\text{DCFT}}\\
                          &\quad\approx  \left\langle\,\prod_{\alpha=1}^{n}\,\left[\CO_{\delta_{\alpha}/2,J_{\alpha}/2}(x_{\alpha},z_{\alpha})\,\CO_{\delta_{\alpha}/2,J_{\alpha}/2} (\bar{x}_{\alpha},\bar z_{\alpha})\right]\,\right\rangle_{\text{CFT}} \ ,
                        \end{aligned} \tag{$\mathbf{S}^\prime$}
 \end{align}
 \normalsize
where $\bar{z}=(\hat{z},-z_\perp)$.
The proof of \eqref{statement spinning} proceeds in parallel with the scalar case \eqref{statement}.
Here we content ourselves with working out the simplest example and relegate the proof to the sequel \cite{long}.

\paragraph{Bulk spin-$J$ one-point function.}
The conformal symmetry determines the bulk one-point function of a spin-$J$ operator in DCFT as:
\small\begin{align}\label{eq:1pt_DCFT spinning}
    \langle\, \slO_{\delta,J} (x,z)\, \rangle_{\text{DCFT}}
    =
    a_{[\delta,J]}\,\frac{\left[(x_\perp\cdot z_\perp)^2-|z_\perp|^2\,|x_\perp|^2\right]^{J/2}}{|x_\perp|^{\delta+J}} \ ,
\end{align}\normalsize
where $x_\perp\cdot z_\perp\equiv x^{i}_{\perp}z_{\perp,i}$
This is indeed kinematically equivalent to the following two-point function in CFT:
\small\begin{align}\label{eq:2pt_aCFT spinning}
\begin{aligned}
    \langle\,&\CO_{\delta/2,J/2} \left(x,z\right)\, \CO_{\delta/2,J/2} \left(\bar x,\bar z\right)\,\rangle_{\text{CFT}}\\
&    =
   C_{[\delta/2,J/2],[\delta/2,J/2]}\\
   &\quad \times\frac{\{(z\cdot\bar{z})\,(x - \bar x)^2-2[z\cdot (x - \bar x) ][\bar{z}\cdot (x - \bar x) ]\}^{J/2}}{|x - \bar x|^{\delta+J}}\\
&    =C_{[\delta/2,J/2],[\delta/2,J/2]}\,
    \frac{\left[(x_\perp\cdot z_\perp)^2-|z_\perp|^2\,|x_\perp|^2\right]^{J/2}}{2^{\delta-J/2}\,|x_\perp|^{\delta+J}}\ .
\end{aligned}
\end{align}\normalsize

\bigskip
\paragraph{Discussion and future direction.}
Correlation functions in DCFTs have attracted much attention in recent studies, and have been explored extensively so far \cite{Billo:2016cpy,Gadde:2016fbj,Kobayashi:2018okw,Guha:2018snh}.
In this letter, we proposed the novel method that can reproduce correlation functions in DCFT by those in the ancillary CFT.
Our method is a complementary approach to the existing method for DCFT correlators.
It is free from additional complications for handling defects, and we believe our approach is more accessible to those who are interested in DCFTs but only familiar with conventional CFTs.

For comparison to ours, let us sketch the previous works \cite{Billo:2016cpy,Gadde:2016fbj} which developed systematic methods to build conformal invariants in DCFT based on the embedding space formalism.
To implement the breaking of conformal symmetry to the subgroup that remains in DCFT, these methods introduce either auxiliary vectors \cite{Gadde:2016fbj} or partial inner products for the embedding space vectors \cite{Billo:2016cpy}.
While the two methods are different-looking at first sight, they are equivalent and can be used interchangeably (see, e.g.,\,\cite{Kobayashi:2018okw}).
Hence, we focus on the latter description in the following.
Then, the construction of the DCFT correlators amounts to enumerating conformal invariants built out of both ordinary and partial inner products of the embedding vectors.
This procedure can be carried out in the same way as the CFT correlators \cite{Costa:2011mg}, but more invariants arise from the partial inner products \cite{Billo:2016cpy}.
There is another possible complication that some of the invariants are not independent due to the lack of full conformal symmetry.
This obstacle can be circumvented in our method which reduces the construction to the mundane problem in CFT.
On the other hand, ours has a drawback that the number of operators is doubled and higher-point correlators have to be dealt with.
It depends on one's familiarity which method is favorable.
DCFT oriented readers may prefer the existing method while others would find ours more intuitive and useful.

While we have mainly focused on bulk spinning local primaries in this paper, one can consider spinning defect local primaries carrying two types of spins associated with the parallel and transverse rotational groups. 
In \cite{long}, we will extend the method of images to include defect local primaries with the parallel and transverse spins and examine the correlation functions with these operators systematically in terms of CFT with the hope of uncovering new structures and relations between correlation functions in DCFT.

Some examples of lower-point functions [from Eq. \eqref{eq:1pt_DCFT} to \eqref{eq:3pt aCFT}] imply that the one-point coefficient $a_{\delta}$ and the bulk-to-defect coupling constant $b_{\delta,\hat{\delta}}$ in DCFT correspond to the two-point $C_{\delta/2,\delta/2}$ and the three-point $C_{\delta/2,\delta/2,\hat{\delta}}$ coefficients in the ancillary CFT, respectively, as
\begin{align}
 C_{\delta/2,\delta/2}= 2^\delta\,  a_{\delta}\ ,\qquad C_{\delta/2,\delta/2,\hat{\delta}}=2^{\delta-\hat{\delta}}\, b_{\delta,\hat{\delta}}\ .
\end{align}
These coefficients incorporate model-dependent information that cannot be determined kinematically by the conformal symmetry. 
This non-kinematical correspondence is not limited to lower-point functions.
One can expand higher-point correlation functions on both sides of \eqref{statement} in a particular basis given by conformally invariant structures associated with its global symmetry and match the model-dependent data of the two theories.
Given the dictionary between the data beyond kinematics in the two theories, the method of images may shed light on new aspects of DCFTs from the viewpoint of conventional CFTs.
For example, it will allow us to constrain the DCFT data by solving the conformal bootstrap equations in DCFT, not directly, but by reducing them to the corresponding equations in CFT and leveraging the well-developed techniques in the past fifteen years since \cite{Rattazzi:2008pe}.
While the DCFT crossing equations do not necessarily have the positive coefficients for the conformal block expansions in the bulk channel, one can perform the conformal bootstrap by assuming solutions with positive coefficients as in \cite{Liendo:2012hy}.
Within such a restricted parameter space, combining the conformal bootstrap with the method of images would be favorable to search for a new class of nontrivial DCFT models.

One of the most fundamental issues in DCFTs is whether an energy condition exists or not.
A promising candidate is the averaged null energy condition (ANEC), which has been shown to hold in any relativistic unitary quantum field theory \cite{Faulkner:2016mzt,Hartman:2016lgu}.
ANEC has played a principal role in restricting central charges of CFTs \cite{Hofman:2008ar}, and it imposes new constraints for the types of defects if it holds in DCFTs (see, e.g., \cite{Jensen:2018rxu,Herzog:2020bqw,Chalabi:2021jud}).
In the presence of extended objects, the existing proof of ANEC is no longer applicable, but even in such a situation, our method would be beneficial to the proof as it boils down to a statement of a CFT correlation function.

In our prescription \eqref{statement spinning}, we used the encoding polynomial techniques to handle the spin degrees of freedom as homogeneity of the polynomial in the polarization vector $z^\mu$.
In a recent study \cite{Kravchuk:2018htv}, this flexibility has been exploited to extend the notion of spin to a continuous variable, i.e., continuous spin. (See also \cite{Caron-Huot:2017vep,Simmons-Duffin:2017nub} for related works.)
One realization of continuous spin is the light-ray operators that can be defined only on Minkowski spacetime.
 Such operators have attracted much attention recently due to the close relation to the ANEC and its higher spin generalizations (see, e.g., \cite{Hofman:2008ar,Hartman:2016lgu,Kravchuk:2018htv,Kundu:2020gkz}). Throughout this paper, we have restricted ourselves to correlators on Euclidean space, and it would also be interesting to extend the method of images to Minkowski spacetime. If there exists a continuous spin operator in DCFT, their CFT counterparts should be a pair of continuous spin operators too. Hence, the method of images turns a correlation function including a ``light-ray operator'' in DCFT into the one with two light-ray operators on the CFT side, whose structure might be revealed by the recent studies of \cite{Kologlu:2019mfz,Kologlu:2019bco,Chang:2020qpj,Chang:2022ryc}.
We leave these intriguing problems for future works.

\bigskip
\begin{acknowledgments}
\paragraph*{Acknowledgments.}
We are grateful to Chris Herzog for his useful comments on the manuscript. We also thank Kohei Fukai for valuable discussions.
The work of T.\,N. was supported in part by the JSPS Grant-in-Aid for Scientific Research (C) No.19K03863, Grant-in-Aid for Scientific Research (A) No.\,21H04469, and
Grant-in-Aid for Transformative Research Areas (A) ``Extreme Universe''
No.\,21H05182 and No.\,21H05190.
The work of Y.\,O. was supported by Forefront Physics and Mathematics Program to Drive Transformation (FoPM), a World-leading Innovative Graduate Study (WINGS) Program, the University of Tokyo.
The work of Y.\,O. was also supported by JSPS fellowship for young students No.\,21J20750, MEXT, and by JSR fellowship, the University of Tokyo.
\end{acknowledgments}

\bibliography{DCFT}

\end{document}